# Virtual spectrometer for radicals' vibrations. 1. Polyacenes and fullerenes

Elena F. Sheka and Vera A. Popova

*Peoples' Friendship University of Russia (RUDN University), 117198 Moscow, Russia*

sheka@icp.ac.ru

**Abstract.** A virtual vibrational spectrometer, based on efficient software exploring unrestricted Hartree-Fock approximation, has been proposed to perform computational spectroscopy of radicals. The computational 'device' allows obtaining spectra of IR absorption and Raman scattering of large molecules. A number of test cases involving naphthalene and pentacene as well as fullerenes $C_{60}$ and $C_{70}$ will be discussed in order to point out strengths, limitations, and ongoing developments of the spectrometer.

## 1. Introduction

For the last two decades, computational vibrational spectroscopy has undergone extensive development, becoming a powerful tool for studying electronic and nuclear dynamics of large molecules [1-6]. The prosperous implementation of the electronic and vibrational theory of molecules in efficient modern computational programs made it possible to provide a reliable concordance of calculated and experimental data. Both spectroscopies address the object on the same atom-structural language making *in vitro* and *in silico* experiments significant independently of each other while suggesting important nuances of the vision of the underlying physicochemical features for both. This trend did not go unnoticed, and in 2012, the group of V. Barone came up with the idea of a virtual multi-frequency spectrometer (VMS) that implements all the power of the *in silico* computational spectroscopy [7]. Particularly important, the suggested VMS provides access to the latest developments beyond a standard rigid-rotor harmonic-oscillator model with application to both vibrational and electronic spectra. The experience of several years of work with such a 'device' has shown impressive results of how complex aspects of electronic and vibrational states, as well as electronic-vibrational interactions, can be revealed in *in silico* experiments, which greatly facilitates the prediction of molecular properties along with their spectroscopic signatures [8]. Unfortunately, so far molecular objects, which can be treated by VMS, have been quite small since the discussed VMS bears the imprint of the capabilities and disadvantages of modern methods of quantum modeling, in connection with which many important objects remain outside the capabilities of the 'device'. This is not surprising, since, usually, the first 'device models' only indicate the horizons of new class of instruments, the achievement of which will take time and efforts of many researchers.

Versatile large stable molecular radicals are such difficult subject of computational spectroscopy. Particularly interesting is an extensive family of $sp^2$ nanocarbons, involving higher polyacenes [9] and polycyclic aromatic hydrocarbons (PAHs) [10, 11], fullerenes [12] and carbon nanotubes [13, 14], as well as a large variety of graphene molecules [15]. The species are on the top of scientific and technological importance for which *in vitro* vibrational spectroscopy is one of the most required analytical tools. However, the relationship between the spectroscopic outcome and the underlying physicochemical properties is often indirect, while the disentanglement of the different factors playing a role in determining the overall result can be strongly facilitated by quantum mechanical (QM) computations. Evidently, a virtual spectrometer similar to VSM is highly required. Nevertheless, the VMS of the Barone's team is not suitable in the case of the above mentioned polyatomic objects due to two reasons. (i). The model molecules consist of tens-and-hundreds of atoms while only small-to-medium-size systems can be nowadays studied by QM methods with an accuracy rivaling that of the most sophisticated experimental techniques [2, 3]. (ii) All radical molecules are open-shell systems for which both VMS conceptual approaches CCSD(T) and DFT are not applicable due to small size of the systems to be analyzed by the former as well as principal obstacles preventing from the consideration of such open-shell systems by the latter (see profound reviews [16-18] and references therein). At the same time, as shown by a large pool of *in silico* experiments based on unrestricted Hartree-Fock (UHF) approach, the latter provides time-saving simulations of large open-shell molecules with the accuracy of the structure determination comparable with experimental data (see a large collection of *in silico* results in [19, 20] and references therein). Moreover, in all the cases when the UHF predictions of various features of molecular radicals could be obtained experimentally, a perfect fitting of *in vitro* and *in silico* data was obtained. Particularly impressive are UHF results concerning trends in the molecular radical behavior. One of the most important conditions for the success achieved in describing the spin-dependent properties of the ground electronic state of fullerenes [19] and graphene molecules [20] was the use of a semi-empirical UHF approach, which made it possible to carry out a huge array of virtual experiments and work with confidence with trends.

The theory of vibrations did not disregard molecular radicals. So back in 1977 Pulai [21] proposed using the UHF approximation to calculate the vibrational spectra of radicals. He with colleagues successfully applied UHF approach for the consideration of vibrational problem of hexatryinyl ($C_6H$) [22] and phenoxyl ($C_6H_5O$) [23] radicals and showed that UHF approach can be effective inexpensive alternative to the MCSCF method for highly correlated organic molecules[1]. With the development of quantum-chemical methods for solving the problem of correlation interaction (CI), the dynamic problem of the radical became the problem of the CI methods, such as, for example, CCSD(T) in the Barone's spectrometer [7, 8]. However, the practical use of these methods is limited by the prohibitively high computational time required for large molecules. The situation changed dramatically when DFT came to the virtual vibrational spectroscopy. Now, the presence of an extensive set of functionals and basis functions makes it possible to select a particle basis-functional pair suitable for a good description of vibrational spectra of practically any class of molecules. Having become dominant for *in silico* spectroscopy, DFT in the form of its UDFT version extended its influence to radicals. However, as shown by numerous calculations [8], the use of UDFT versions did not reveal significant deviations in the dynamic solutions obtained using the conventional DFT one, as a result of which there was a tacit agreement that the common DFT approach is quite sufficient for calculating the spectra of radicals. And numerous publications on the calculation of vibrational spectra of polyacenes, polyaromatic hydrocarbons, fullerenes (see [25-34] but a few) poured down like a cornucopia. Therefore, the dynamics of all this radical wealth is presented from the thousands of pages of publications as the dynamics of closed-shell molecules.

---

[1] Twenty years later phenoxy radical again became the object of comparative testing of a set of computational methods for open-shells systems, including DFT, ROHF, UHF and general complex Hartree-Fock (GCHF) approach [24]. Again, UHF showed its high efficiency and practical identity according to the results obtained by the method. UHF has defended its right to describe electronic states of radicals quite correctly [18].

Collective confidence in the sufficiency of the correct selection of a basis-functional pair and the use of any publicly available DFT program to obtain a satisfactory description of the experimental spectrum of any molecule, including the radical, raises two questions:

1. Is the vibrational spectrum of a radical really insensitive to the spin features of the electronic state of the molecule?

2. Isn't this insensitivity of the DFT vibrational spectrum a consequence of that for the DFT inapplicability to the ground state of the radical?

To answer these questions, we propose to use a specific UHF virtual vibrational spectrometer (UHF-VVS) to carry out an extensive computational experiment on a large set of radicals belonging to a single family of $sp^2$ nanocarbons, at the same level of the theory. The approach high efficacy allows forming the ground for an extensive comparative study for the main trends of the $sp^2$ radicals dynamic features to be revealed. The current paper presents Part 1 of the study and concerns the grounds of the proposed spectrometer and first results concerning vibrational spectra of benzene, naphthalene, pentacene as well as fullerenes $C_{60}$ and $C_{70}$. All virtual results are inspected with respect to experimental data. Part 2 is tightly bound with the recent *in vitro* vibrational spectroscopy results related to 'graphene signatures' of $sp^2$ amorphous carbons [35-37] and will concern graphene molecules [38].

## 2. UHF virtual vibrational spectrometer

The spectrometer implements a standard rigid-rotor harmonic-oscillator model in the framework of semi empirical UHF QM approach (either AM1 or PM3) implemented in the software CLUSTER-Z1 [39, 40]. The codes laid the foundation for a big number of *in silico* experiments performed with large open-shell systems in a perfect concordance with experimental data concerning equilibrium structure and electronic properties of the species in the ground state (see numerous examples in [19, 20] and references therein). The vibrational part of the codes concerns the calculation of one-phonon IR absorption and Raman scattering spectra based on molecular structure, either optimized or input from outside. Analytic force constants and first derivatives of electric and magnetic properties are calculated following the Pulay's analytical determination of the total energy gradients [21, 41, 42]. The intensity of vibration bands are determined by the first derivatives of dipole moments and polarizibility fensos over normal coordinates for the light absorption and scattering, respectively. Extended codes NANOVIBR [43] were developed to perform parallel HF computing of vibrational spectra of molecules consisting of more than 1000 atoms. The code provides performing calculations in the restricted Hartree-Fock (RHF) as well thus allowing a comparative RHF-UHF investigation of the molecular dynamics under the same conditions.

However, it is well known that the harmonic-oscillator approximation usually overestimate force constant matrix elements due to, mainly, not taking into account correlation of the molecule electrons and anharmonicity of the motion of its nuclei The latter leads, in particular, to a truncated description of the vibrational spectra due to vanishing intensities for overtones and combination bands. These shortcomings were taken into account in the VMS discussed above, which provides the calculation of vibrational spectra at the CCSD(T) level involving the correlation of electrons and exploring all the latest developments in the theory of anharmonicity and its computational implementation. The VMS application has convincingly shown [8] that it is mandatory to go beyond the above double-harmonic approximation accounting for both mechanical and electrical anharmonic effects [44] in order to fulfill the accuracy (for frequencies) and interpretability (for intensities) needs of spectroscopists. Unfortunately, numerous attempts to realize the requirements occurred of high computational cost (see for a review [45] and references therein). Among the alternatives, the most successful ones are those based on the vibrational self-consistent field or vibrational second-order perturbation theory (VPT2). While the calculation of vibrational energies at the VPT2 level is now a routine task widespread in various computational packages, work on transition intensities has been scarcer. The best balance was achieved in the considered VMS.

The software package CLUSTER-Z1 partially copes with the first problem, since the UHF algorithm makes it possible to take into account the correlation of electrons in large systems. However, the lack of anharmonic corrections calculation dooms the user of UHF-VVS to deliberately overestimated values of vibrational frequencies and serious problems with determining the intensities in the calculated spectra. The methods of *post factum* scaling developed to date make it possible to cope with the overestimation of frequencies [46, 47]. However, the second problem, concerning the intensity of the bands, becomes possible only as a result of the development of algorithms for solving anharmonic problems, which is perfectly demonstrated by examples of using the spectrometer. The aim of this study is to show that, despite these obvious disadvantages of the spectrometer based on the CLUSTER-Z1 package, the calculated spectra obtained with its help reveal subtle features of the spectra caused by the spin nature of the ground state of radical molecules. These effects, in turn, allow a new look at the experimental spectra of such molecules and propose new aspects of the interpretation of the latter.

All calculations, discussed in Part 1 and 2 of the current study, were performed by using AM1 version of CLUSTER-Z1. No scaling method was applied for the comparison of RHF and UHF results to be undisturbed. Experimental data were taken from the literature. Only *sp²* carbonaceous species were considered. All the species are characterized by odd electrons, interaction with which provides either the formation a covalent bond at low interatomic distance $R$ up to critical value $R_{cov}$ or is not enough to complete the bonding when $R > R_{cov}$ leaving the latter radicalized [49]. Radicalization of the studied molecules is characterized by a specific set of identifying criterion parameters involving the energy shift $\Delta E_{RU} = E_{RHF} - E_{UHF}$; the number of effectively unpaired electrons $N_D$; and spin contamination $\Delta \hat{S}^2$, $S$ is total spin. The detailed description of the quantities can be found elsewhere [18-20]. All the values are provided with the UHF-VVS at the stage of the molecule structure optimization and their nonzero quantity exhibit a radical nature of the molecule. Table 1 lists the data related to the molecules of the study Part 1.

3. I*n silico* vibrational spectra of polyacenes

For the first stage of testing the efficiency of the UHF-VVS, three molecules were selected: benzene, naphthalene, and pentacene, which are the most studied spectral-dynamic representatives of the polyacene family. The history of the *in silico* spectroscopy of each of these molecules, considered as closed-shell electronic structures, is very rich and contains a large number of virtual results using many different methods aimed at achieving the best fitting between the calculated data and experimental spectra. However, not the competition of open-shell results for the best fitting, provided with the spectrometer, but the identification of new trends caused by the spin features of the electronic system in the ground state is the purpose of the first testing of the spectrometer. Because of this, everywhere below, comparing the obtained open-shell spectra with the closed-shell ones, we will restrict ourselves only to the RHF-UHF pairs obtained at the same level of theory. The practical significance of this pair of spectra will be analyzed by their comparison with experimental data.

**Table 1**. Identifying criterion parameters of the odd electron correlation in open-shell $sp^2$ molecules*

| Molecule | Spin multiplicity, $\Delta N_{\alpha\beta}$** | Odd electrons $N_{odd}$ | $\Delta E_{RU}$*** kcal/mol | $N_D$, e⁻ | $\Delta \hat{S}^2_U$ |
|---|---|---|---|---|---|
| benzene | singlet | 6 | 0 | 0 | 0 |
| naphthalene | singlet | 10 | 1.85 | 1.47 | 0.74 |
| pentacene | singlet | 22 | 24.24 | 5.54 | 2.77 |
| fullerene C$_{60}$ | singlet | 60 | 17.26 | 9.87 | 4.94 |
| fullerene C$_{70}$ | singlet | 70 | 40.80 | 14.38 | 7.19 |

\* Criteria of odd electrons correlation
*Criterion 1*:
$$\Delta E_{RU} \geq 0, \text{ where } \Delta E_{RU} = E_{RHF} - E_{UHF}$$
presents a misalignment of energy. Here, $E_{RHF}$ and $E_{UHF}$ are total energies calculated by using restricted and unrestricted versions of the program in use.
*Criterion 2*:
$N_D \neq 0$, where $N_D$ is the total number of effectively unpaired electrons and is determined as
$$N_D = trD(r|r') \neq 0 \text{ and } N_D = \sum_A D_A.$$
Here, $D(r|r')$ and $D_A$ present the total and atom-fractioned spin density caused by the spin asymmetry due to the location of electrons with different spins in different spaces.
*Criterion 3*:
$$\Delta \hat{S}^2 \geq 0, \text{ where } \Delta \hat{S}^2 = \hat{S}^2_U - S(S+1)$$
presents the misalignment of squared spin. Here, $\hat{S}^2_U$ is the squared spin calculated within the applied unrestricted technique while *S(S+1)* presents the exact value of $\hat{S}^2$.

Criterion 1 follows from a well known fact that the electron correlation, if available, lowers the total energy. Criterion 2 highlights the fact that the electron correlation is accompanied with the appearance of effectively unpaired electrons that provide the molecule radicalization. Criterion 3 is the manifestation of the spin contamination of unrestricted single-determinant solutions; the stronger electron correlation, the bigger spin contamination of the studied spin state.

** Formal spin multiplicity is determined by the difference $\Delta N_{\alpha\beta} = N_\alpha - N_\beta$, $N_\alpha$ and $N_\beta$ are the numbers of the molecule electrons with spins α and β, respectively

*** AM1 version of UHF codes of CLUSTER-Z1. Presented energy values are rounded off to an integer.

### 3.1. Benzene

Benzene molecule has been the main object of molecular dynamic problem from the inception of computational vibrational problem (see monograph [49] and references therein). Manually selected forth constants were repeatedly used later on when calculating the vibrational spectra of numerous benzene derivatives (see [50, 51] and references therein). Development of quantum chemistry has radically changed the situation. Benzene was the first molecule on which P.Pulay assayed the quantum chemical concept of computational molecular vibrational dynamics [52, 53]. The molecule has been a platzdarm for numerous theoretical investigation since as new computational programs appear, allowing vibrational spectrum of such a molecule to be calculated. One of the last attempt was performed by using DMOL$^3$ program [54, 55]. Anharmonic effects were carefully studied for the molecule as well [56]. Experimentally, IR and Raman spectra of the molecule were obtained in gaseous, liquid, and solid states.

As follows from Table 1, benzene is not a radical. This is due to the length of the molecule C=C bonds equal to 1.395 Å, that is just a critical $R_{cov}$ for the *sp$^2$* species before the bonds radicalization starts [57]. Naturally, virtual RHF and UHF spectra are identical due to which the only question to the UHF-VVS testing is how well Hartree-Fock harmonic approximation provides the reproduction of experimental spectra of the molecule. Red and blue plottings in Fig. 1 present experimental and UHF virtual spectra of the molecule. Experimental data (Raman scattering [58] and IR absorption [59]) are related to free molecules in the gaseous state.

Two comments should be discussed before analyzing the spectra observed, which is important not only for benzene, but for other polyacene molecules as well. Both concern the facilitation of the spectra description. The first concerns usual approach to the experimental spectra interpretation from the position of standard *group frequencies* (GFs) [60, 61]. Those related to hydrocarbon aromatic molecules are listed in Table 2. The next issue concerns the evident difference of Raman and IR spectra, which is typical to not only benzene, but also other *sp$^2$* hydrocarbons. As occurred, the feature is connected with the type of covalent bonds participating in each of vibrational mode. Thus, vibrational modes are IR-active if there is a change in dipole moment associated with vibrations. Covalent homopolar bonds of molecules do not normally possess a static dipole moment due to which their activity arises from a dynamic mechanism in which the charge is transferred from extended to compressed bonds [62]. Concerning homopolar C=C bonds with nil static moments, IR intensity depends on the charge value, which for a predominant majority of carbon atoms is small. Consequently, vibrational modes related to carbon atoms are expected to be practically IR-inactive. Therefore, IR spectroscopy of *sp$^2$* hydrocarbons is a mean to mainly study heteroatom framing of their benzenoid hexagons with the strongest preference of C-H bonds characterized by the largest static moments. In contrast, the polarizability is quite favorable for homopolar bands and because these bonds dominate in the species, Raman scattering becomes one of the main techniques to study the structure of their carbon cores. This conclusion is generally supported by the pool of available spectroscopic data with the only exclusion concerning high frequency C-H(R) stretchings [61]. Perhaps, a pretty similar behavior of the bonds in both aliphatic and aromatic compounds is responsible for the feature. The same can be said about the spectra of benzene. Herein, all the wavenumbered attributions will be given according experimental spectra. As seen in Fig. 1, both experimental and virtual spectral pairs are well similar in the region of C-H stretchings over 3000 cm$^{-1}$, while drastically different in the region below 2000 cm$^{-1}$. In this region, the main features are attributed to the benzene ring breathing in Raman spectrum and to out-of-plane in phase deformational vibrations of C-H bonds in the IR spectrum. Both conclusions are supported by quantum-chemical calculations in the benzene case [58, 63].

Virtual spectra in Fig. 1 reproduce the experimental counterparts quite well with respect to the main spectral features. The expected upshifting of the former is large and constitutes ~100 – 150 cm$^{-1}$ in average. The value is consistent with anharmonic corrections [56] related to the C-H stretchings at 3050 cm$^{-1}$ while considerably exceeds the latter at 700 cm$^{-1}$, 1000 cm$^{-1}$, and 1500 cm$^{-1}$. The finding evidences nonuniform total energy dependence from different normal coordinates, thus distinguishing *op* and *ip* in phase δ (C, H) modes of IR spectrum from ring breathing and C=C stretchings observed in the Raman one.

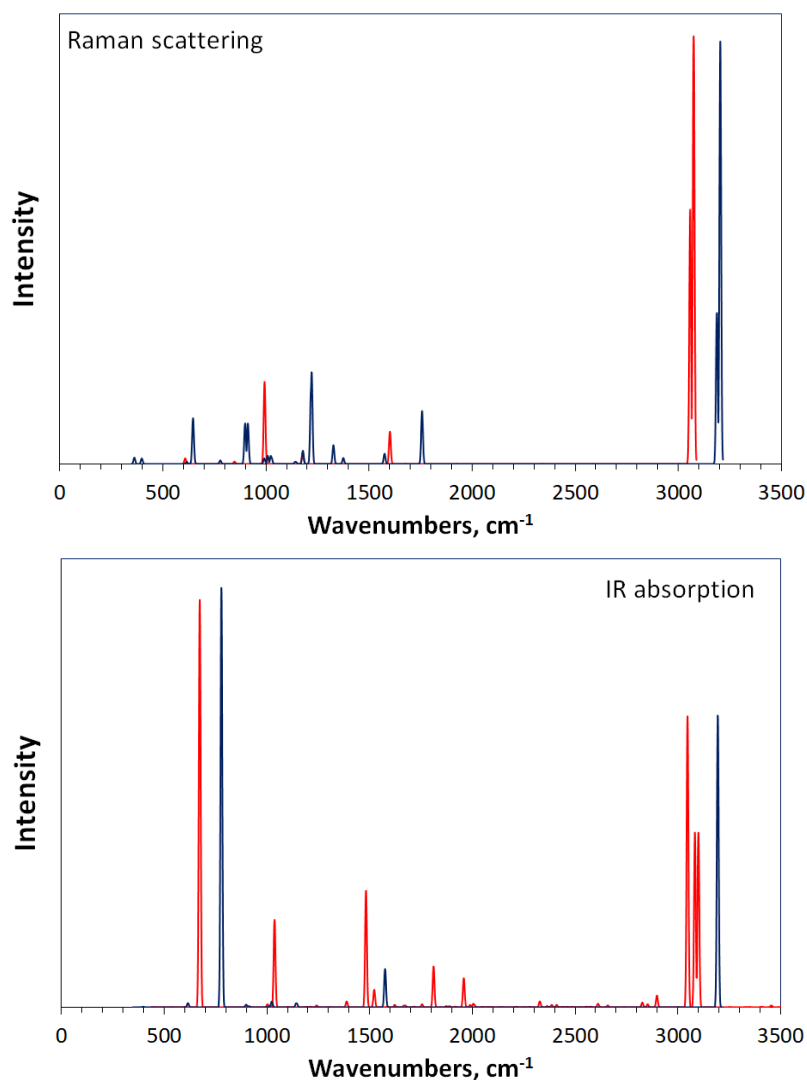

**Figure 1**. Raman and IR experimental (red) and UHF virtual (blue) spectra of benzene. Stick-bar data were convoluted with Gaussian bandwidth of 10 cm$^{-1}$. Intensities are reported in arbitrary units.

### 3.2. Naphthalene

Naphthalene is the second most studied molecule of the polyacene family (see [25-27] and references therein). This molecule was the largest one studied beyond harmonic approximation in the concept of the VMS [45]. In contrast to benzene, characterized by the only C=C bond of 1.395 Å, three kinds of the latter of 1.385 Å, 1.411 Å, and 1.420 Å form the molecule structure. The length of two longer ones exceeds the radicalization criterion $R_{cov} = 1.395$ Å, due to which the molecule becomes radicalized as seen in Table 1. Subjected to the UHF-VVS treatment, the molecule radicalization has become evident in its virtual spectra as well. As seen in Fig. 2, the transition from the RHF to UHF approach complicates Raman and IR differently as a whole and in different spectral regions. Thus, as in the case of benzene, there is no change in the region of C-H stretchings at 3200 cm$^{-1}$. However, below 2000 cm$^{-1}$ both Raman and IR spectra change significantly, Raman spectrum more pronouncedly, as it can be seen in Fig. 3. The difference of the RHF and UHF spectra is not connected with the symmetry, which remains the same $D_{2h}$ in both cases.

First, a few words about experimental spectra. Those are related to microcrystalline (Raman scattering, [27]) and argon-matrix-isolated (IR absorption [26]) samples at room temperature. As seen in Fig. 3, IR spectrum of naphthalene conserves the benzene main feature thus evidencing that *op in phase* (C, H) bendings play the main role in this case as well. In contrast to benzene, the dominating band is accompanied with other modes, the right-hand of which are mainly complex tones caused by anharmonicity [45]. Oppositely to IR spectrum, the Raman one of naphthalene drastically differs from that of benzene. Not breathing ring mode, but C=C stretchings, the number of which increased 2.5 times compared to benzene, take on the role of core participants in the spectrum in the region of 1400-1700 cm⁻¹. Skeletal breathing mode at ~1020 cm⁻¹ and skeletal distortions at ~760 cm⁻¹ and 500 cm⁻¹ complete the formation of the spectrum (the mode assignment is taken from [45]).

**Table 2.** Standard group frequencies of aromatic molecules required for the fractional analysis of vibrational spectra of *sp²* amorphous carbons

| Vibrational frequency, cm⁻¹ | Group frequencies* | |
|---|---|---|
| | (C, C)** | (C, H₁)** |
| 400–700 | 404 δ *op* C-C-C<br>606 δ *ip* C-C-C | – |
| 700–1200 | 707 C-C-C puckering,<br>993 ring breathing,<br>1010 δ C-C-C trigonal | 673 δ *op in phase*<br>846 δ *op*, C₆ libration<br>967 δ *op*<br>990 δ *op*, trigonal<br>1037 δ *ip*<br>1146 δ *ip*, trigonal<br>1178 δ *ip* |
| 1200–1600 | 1309 ν C-C Kekule,<br>1482 ν C-C<br>1599 ν C-C | 1350 δ *ip* in phase |
| 2800–3200 | | 3056 ν<br>3057 ν trigonal<br>3064 ν<br>3073 ν *in phase* |

¹ Greek symbols ν, δ, ρ, *r*, τ mark stretching, bending, rocking, rotational, and torsion modes, respectively, *ip* and *z* mark in plane and out of plane modes.
² Group frequencies notifications of experimental fundamental vibrations of benzene molecule [64].

Both IR RHF and UHF virtual spectra are generally well consistent with experimental one. The situation with the Raman spectrum is more complicated. Leaving aside the differences regarding the frequency values, let us turn our attention to the general structure of the spectrum. The virtual spectra are well coherent with the picture suggested above based on experimental spectrum, albeit quite differently: the fine structure of the RHF spectrum is scarcer than that of experimental, while the structure of the UHF spectrum is obviously richer. Since the fine structure of the UHF spectrum is quite intense (see the opposite situation with the structure of the IR spectrum in Fig. 3), it is possible to suggest that the fine structure in the experimental Raman spectrum may evidence the open-shell nature of the naphthalene ground state. However, the radicalization of naphthalene is quite weak for this conclusion to be affirmative.

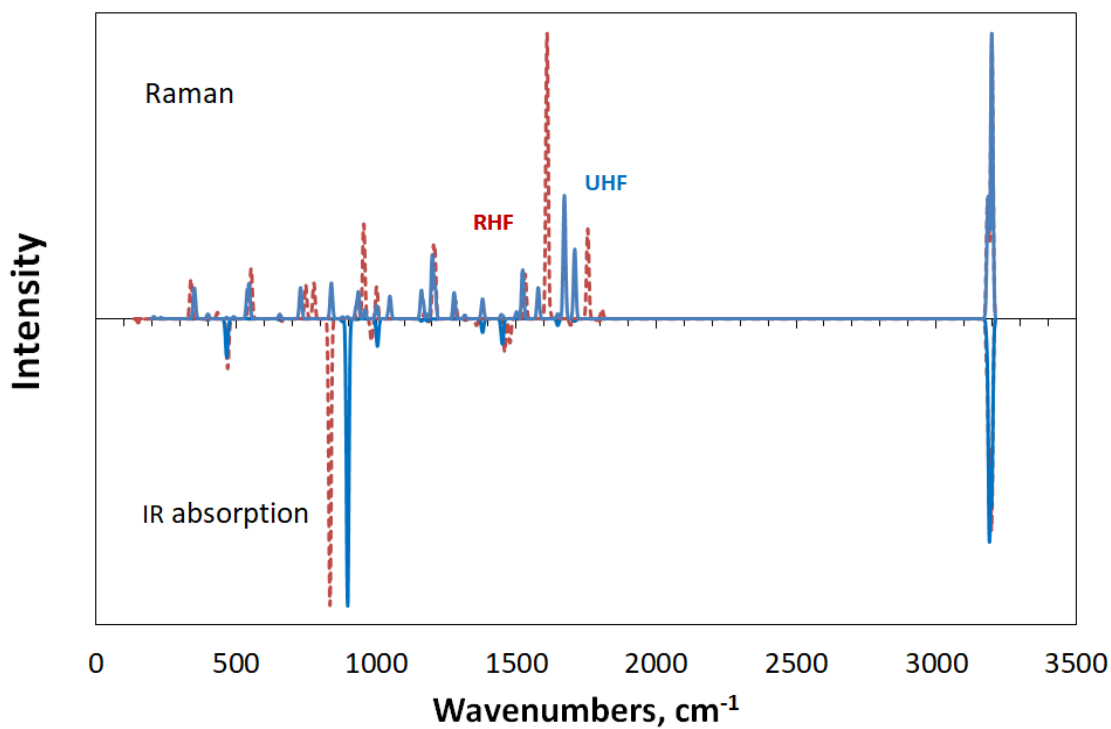

**Figure 2.** Virtual Raman and IR absorption spectra of naphthalene in RHF (dark red) and UHF (bleu) approaches. Stick-bar data were convoluted with Gaussian bandwidth of 10 cm$^{-1}$. Intensities are reported in arbitrary units.

### 3. 3. Pentacene

Among existing linear polyacenes, pentacene ($C_{22}H_{14}$) is the only molecule, radicalization of which is not only predicted, but also directly confirmed experimentally (see [65] and references therein). As seen from Table 1, 25 % of odd electrons of the molecule carbon atoms are effectively unpaired that is why both correlation of odd electrons and the molecule radicalization is quite strong. The molecule symmetry is $D_{2h}$ in both RHF and UHF approaches.

Expectedly, the transition from the RHF to UHF approach influences the molecule vibration spectra markedly. As seen in Fig. 4, practically both Raman and IR spectra are fully reconstructed, which is particularly vivid for the C=C stretchings and bending C-H vibrations in the Raman and IR spectra, respectively. The feature is directly connected with the reconstruction of the C=C bonds pool, which is illustrated in Fig. 5. Attention is drawn to the noticeable richness of the fine structure of both Raman spectra in contrast to a more modest picture for IR spectra. The feature is similar to that of the naphthalene spectra in Fig. 2. Comparing the spectra of the two molecules with that one of benzene in Fig. 1, we see common trends when going from benzene to polyacenes: in the Raman spectra, the ring-breathing mode of benzene transfers its lead role to the C=C stretchings, which become dominating in pentacene. In the IR spectra, the priority of the *out-of-plane in phase* CH bending of benzene is gradually extended to a large set of out-of-plane bendings in the region of 700-1100 cm$^{-1}$ (see Table 2). Attention is drawn to the noticeable richness of the fine structure of both Raman spectra in contrast to a more modest picture for the IR ones.

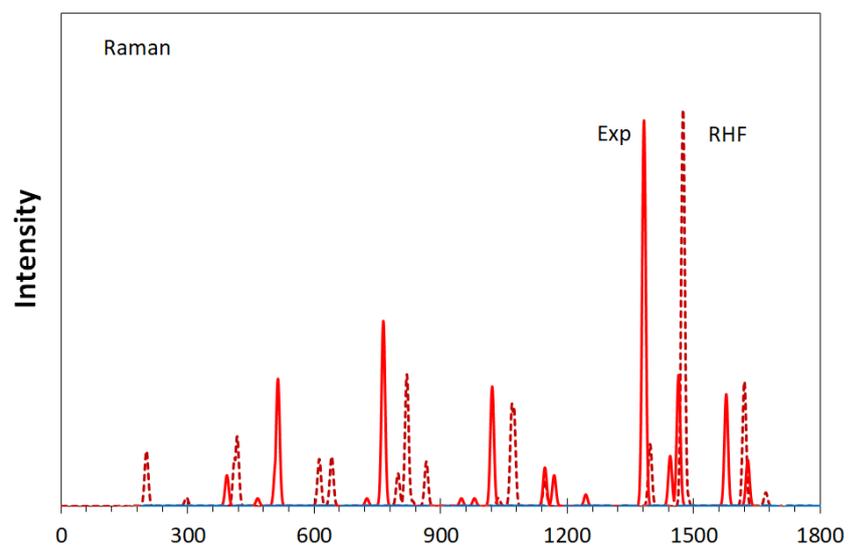
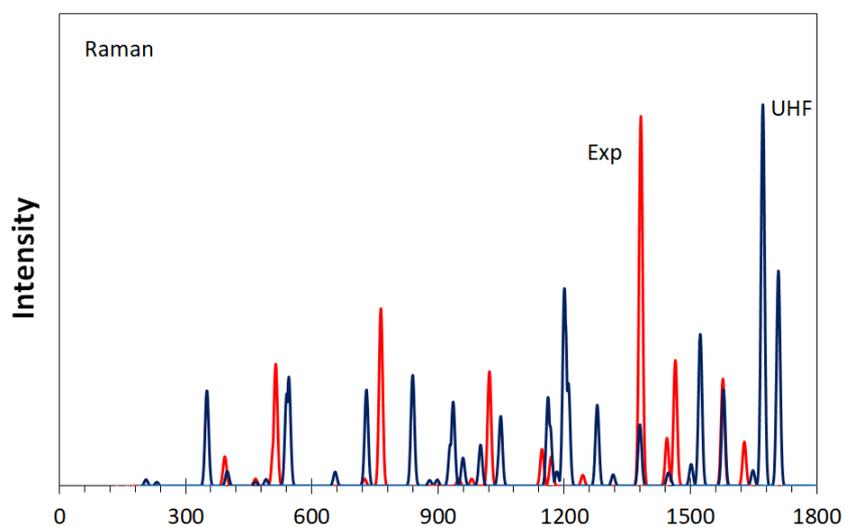
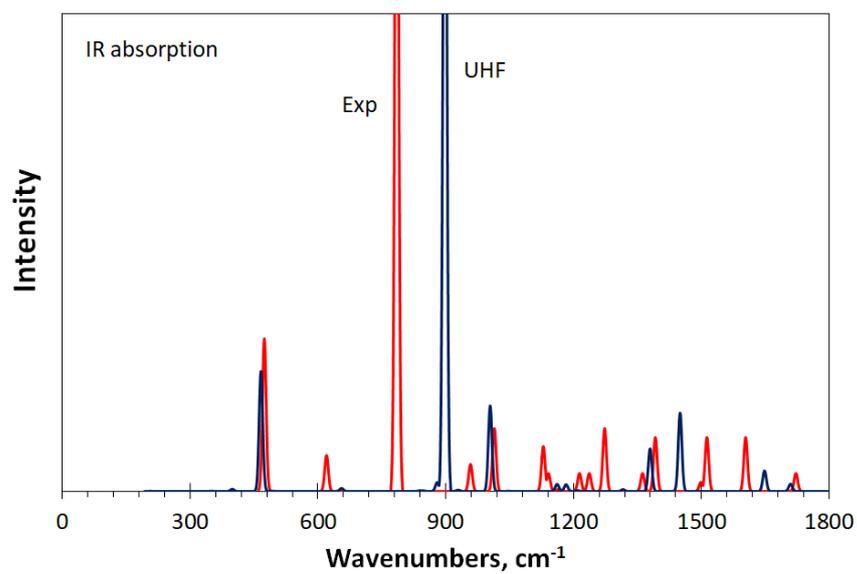

**Figure 3.** Raman and IR experimental (red) as well as RHF (broken dark red) and UHF (blue) virtual spectra of naphthalene. Stick-bar data were convoluted with Gaussian bandwidth of 10 cm$^{-1}$. Intensities are reported in arbitrary units.

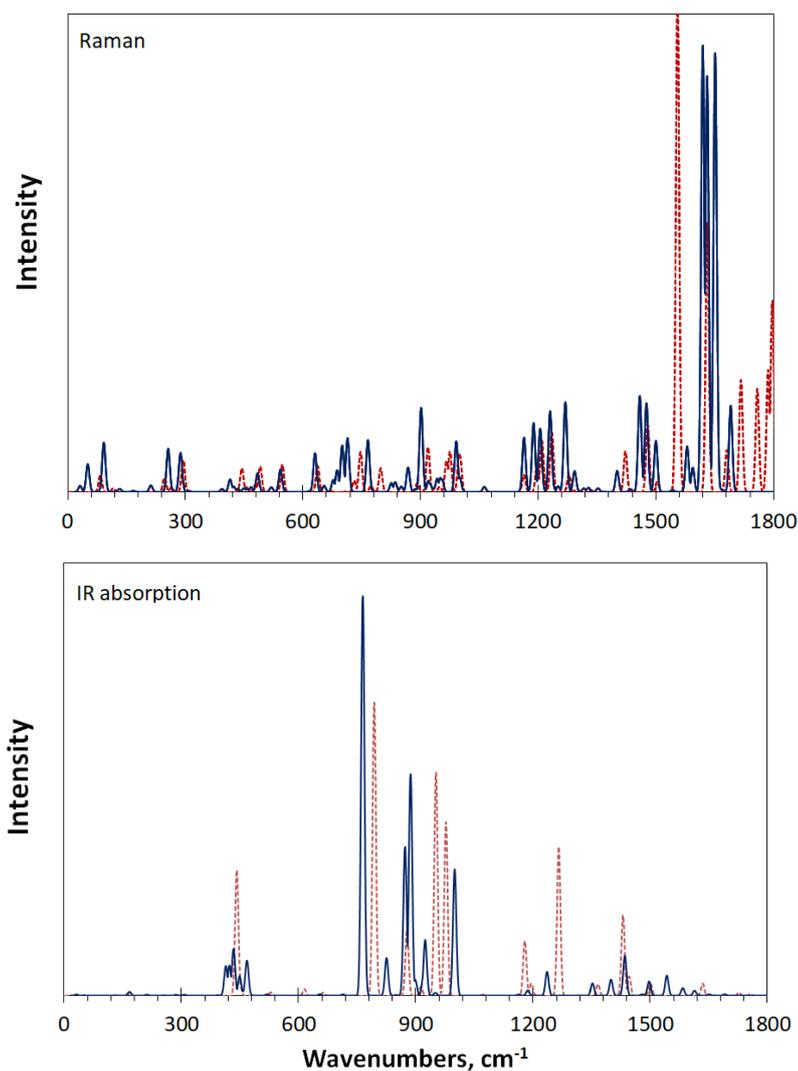

**Figure 4.** Virtual Raman and IR RHF (broken dark red) and UHF (blue) spectra of pentacene. Stick-bar data are convoluted with Gaussian bandwidth of 10 cm$^{-1}$. Intensities are reported in arbitrary units.

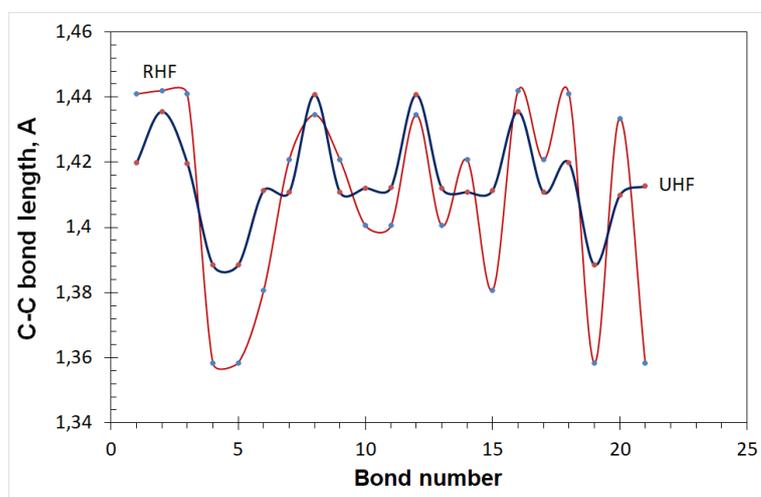

**Figure 5.** C=C bonds length distribution in pentacene

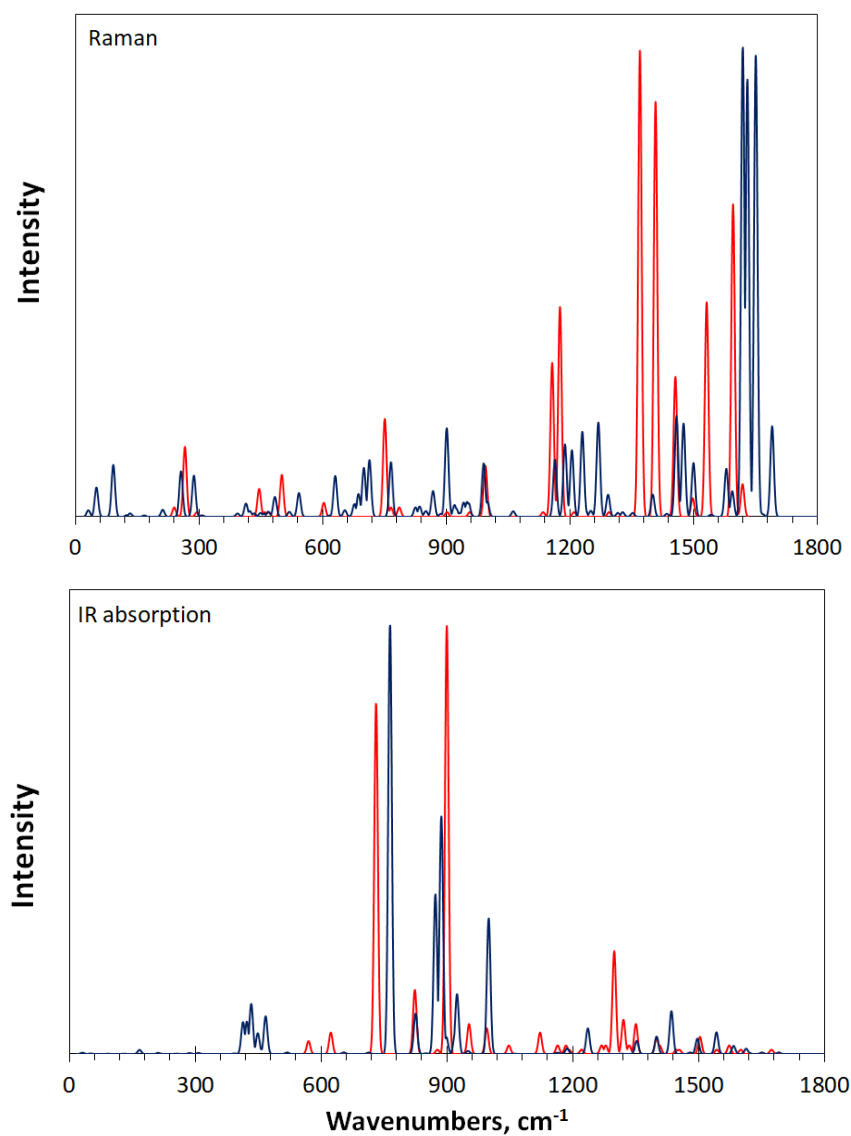

**Figure 6**. Raman and IR experimental (red) and UHF virtual (blue) spectra of pentacene. Stick-bar data are convoluted with Gaussian bandwidth of 10 cm$^{-1}$. Intensities are reported in arbitrary units.

Experimentally, vibrational spectra of pentacene have been carefully studied in matrix-isolated and crystalline states [25-27, 66, 67]. Despite vivid difference of the RHF and UHF virtual spectra, they both reproduce the general picture of experimental spectra discussed above, while a detailed fitting, provided with the UHF spectrum, is much better. Figure 6 displays a comparative view on the Raman and IR pairs of experimental and UHF spectra.

## 4. Fullerenes

### 4.1. Fullerene $C_{60}$

As seen in Table 1, fullerene $C_{60}$ is characterized by high values of the identifying criterion parameters evidencing its radical nature. Virtual UHF spin chemical physics of the substance, convincingly supported with experimental findings, is largely discussed in [19]. UHF virtual vibtational spectroscopy of the species is presented in the current study for the first time.

Radical properties of the molecule are caused by the fact that a large part of C=C distances exceeds the critical value $R_{cov} = 1.395$ Å so that a significant part of effectively unpaired electrons $N_D$ looks quite natural. As shown by virtual structural experiments, transmission from the RHF to UHF approach changes RHF $I_h$ symmetry for UHF $C_i$. This symmetry lowering is not connected with changing either the molecule shape or C=C bond character and is concerned with some delicate quantitative characteristics that concern the scattering of C-C bonds lengths. Figure 7a presents the dispersion of C=C bond distribution in $C_{60}$ structures that correspond to RHF and UBS HF solutions while Figure 7b reveals the bond transformation. However, analyzed from the concept of continuous symmetry, $C_i$ symmetry of the $C_{60}$ molecule is practically $I_h$ [68]. The high continuous symmetry of the molecule provides high-symmetry patterns of all the symmetry sensitive experimental recordings. At the same time, its deviation from the exact $I_h$ symmetry may be used to explain the inconsistencies of experimental recordings from the exact high-symmetry ones. This comment will be used in what follows when analyzing vibrational spectra of the molecule.

RHF and UHF virtual vibrational spectra of fullerene $C_{60}$ are shown in Fig. 8. As seen in the figure, RHF spectra are extremely scarce in a full accordance with the point $I_h$ symmetry. In contrast, high continuous symmetry of the molecule UHF structure calls to life a large number of additional bands in both Raman and IR spectra. Optical spectra, both electronic and vibrational, are usually accepted as empirical arbiters of symmetry problem. In the case of fullerene $C_{60}$, the job occurred the problem since experimental spectra of both types do not confirm an exact point group $I_h$, being much richer in fine structure than predicted by the symmetry [70]. Some peculiarities, such as a weak pure electronic $0_0^0$ transition in both absorption and fluorescence spectra, the vibronic series in the latter spectrum based on symmetry-forbidden *g*-symmetry vibrations, absolutely allowed pattern of the phosphorescence spectrum, and the 'silent' modes in Raman and IR spectra are inconsistent with the high symmetry of the molecule and cast some suspicion on the point. The 'silent' modes have been the stumbling blocks for the interpreters of the molecule vibrational spectra from the very beginning (see exhausted reviews [71-73] and references therein).

The concept of 'silent' modes is tightly connected with computational results on the first-order Raman and IR spectra in harmonic approximation (see a profound review [72]). All the calculations are performed in the closed-shell approximation, according to which both first-order Raman and IR spectra involves only one-two modes (there are generated among them) each thus leaving the remaining modes as 'silent'. Empirically, both spectra are quite rich although their shape varies quite considerably, since all of the existing data on the vibrational properties of $C_{60}$ stem from measurements carried out on thin films, crystals, and solutions. Dresselhauses et al [71] suggested Raman [74, 75] spectra of thin films to be the most appropriate to those of a free molecule. As seen in Figs. 11.10 and 11.12 in [71], these spectra are well similar to the UHF virtual Raman spectrum in Fig. 8. Less similar are experimental IR spectra [76, 77] in Figs. 11.10 and 11.13 in [71] and both RHF and UHF IR spectra in Fig. 8, among which the UHF one is the most similar by overall fine structure. The dissimilarity concerns not the number of the main bands, but their intensity. At the same time, the fact of highly developed fine structure provided with 'silent' modes has been convincingly fixed. Moreover, experiments disclosed high-order spectra in both cases, thus evidencing a strong anharmonicity of the molecule vibrations.

Until now, the presence of 'silent' modes is explained either by disturbing the high symmetry of the molecule or by the second-order effects. The former is usually attributed to either isotopic $C^{12} \rightarrow C^{13}$ substitutions or intermolecular interactions in the solid state. However, the latest studies show [31, 73] that both effects should be unable to explain so rich fine structure of experimental spectra. At the same time, presented in Fig. 8 convincingly demonstrates that highly extended fine structure of both Raman and IR spectra is effect of the first order and is caused by the radical structure of fullerene $C_{60}$, which makes it necessary to use two-determinant HF approach to describe its vibrational spectra.

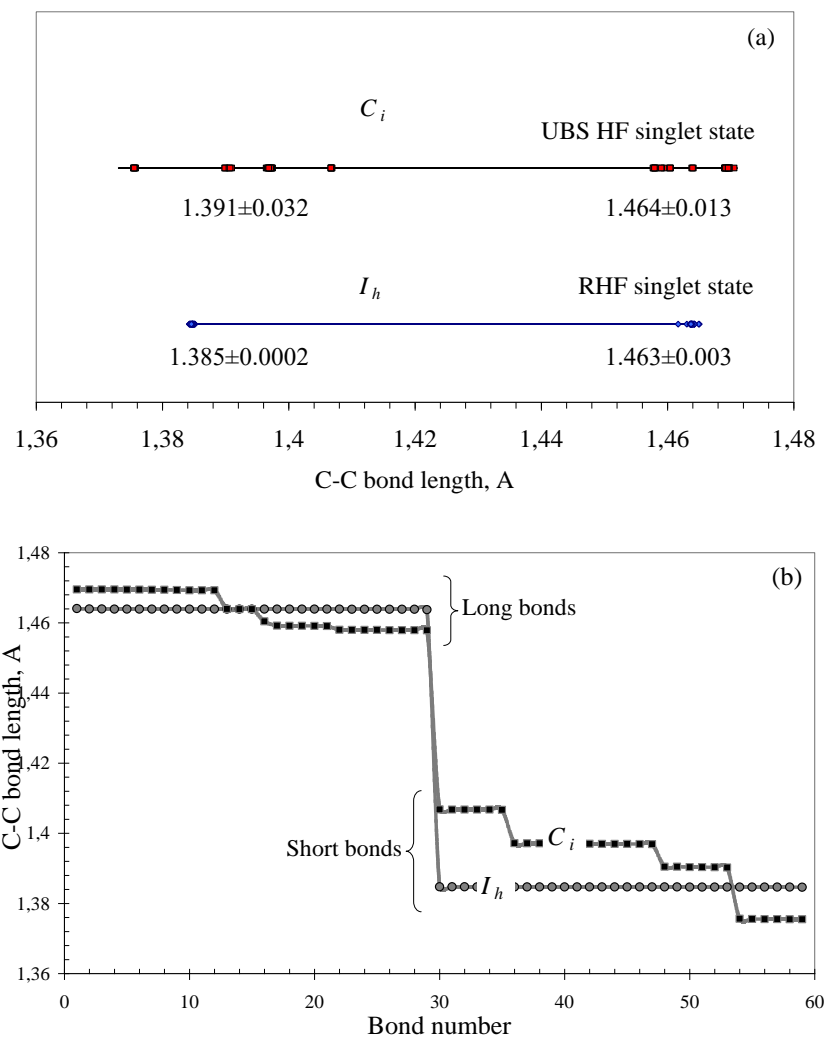

**Figure 7**. Correlation between C=C bonds of $C_{60}$ fullerene related to restricted and unrestricted HF calculations [69]. (a) Bond dispersion; (b) transformation of the Z→A distributions of $sp^2$C-C bonds.

### 4.2. Fullerene $C_{70}$

Identifying criterion parameters in Table 1 clearly reveal a radical property of the molecule. Oppositely to $C_{60}$, changing the HF approach from RHF to UHF does not cause changing in the molecule symmetry, leaving it $D_{5h}$ in both cases. Nevertheless, the pull of C=C bonds undergoes a marked changing which is presented in Fig. 9. The figure presents a redistribution of the bonds over their length thus exhibiting a fundamental reason for changing in the RHF and UHF virtual vibrational spectra to be expected. Responding to the expectations, Fig. 10 presents RHF-UHF pairs of the Raman and IR spectra of the molecule.

Despite the molecule symmetry is conserved under changing the approach, a considerable redistribution of bonds lengths leads to a remarkable changing in the spectra image. Evidently, the Raman spectrum becomes a spectrum of silent-mode-fine structure, in contrast to the silent-mode-free RHF, similarly to the case of $C_{60}$. The feature is less sharp in the case of the IR spectra, since both RHF and UHF spectra are silent-mode-fine ones, albeit with much less extended structure as the UHF Raman one.

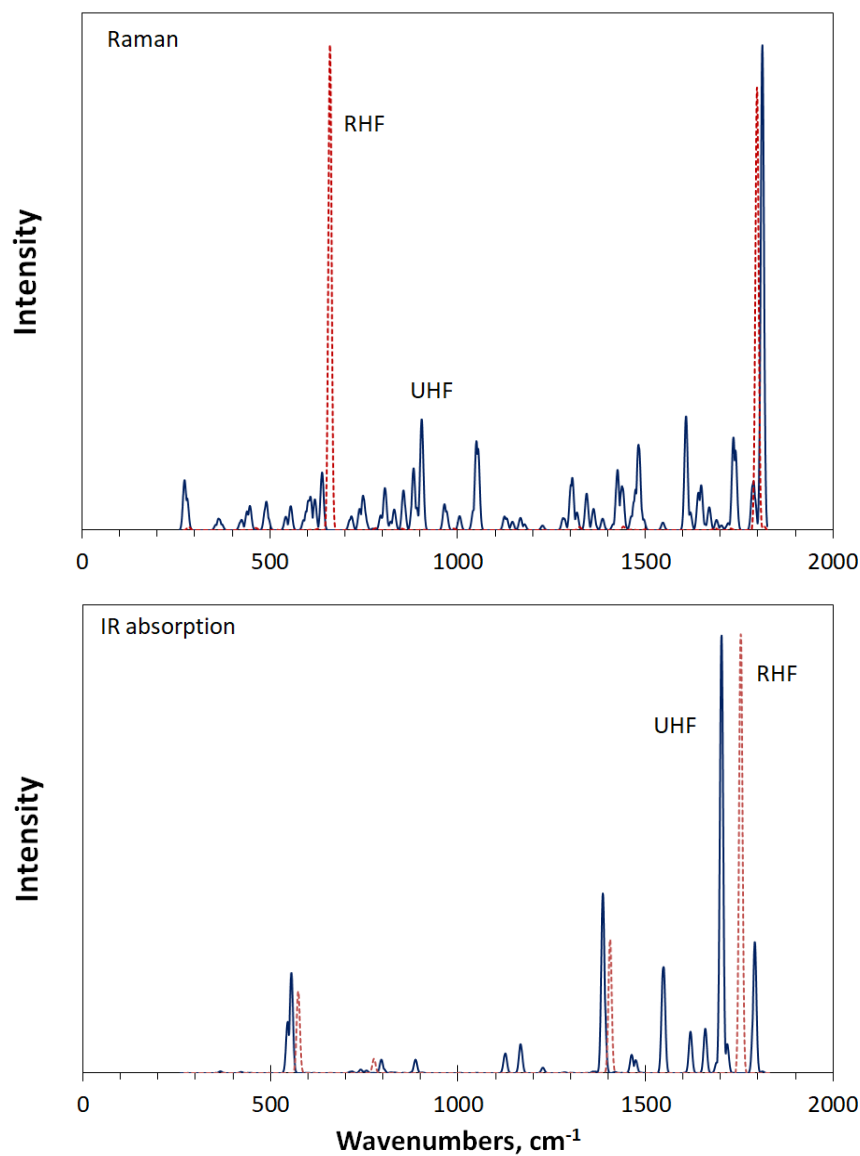

**Figure 8.** Raman and IR RHF (broken dark red) and UHF (blue) virtual spectra of fullerene $C_{60}$. Stick-bar data are convoluted with Gaussian bandwidth of 10 cm$^{-1}$. Intensities are reported in arbitrary units.

Vibrational spectroscopy of $C_{70}$ is not so developed as the $C_{60}$ one [32, 33, 71] apparently due to a big similarity in the behavior of both fullerenes. Due to lower symmetry, the problem of the $C_{70}$ 'silent' modes is not so sharp, as in the case of $C_{60}$. However, closed-shell computational consideration predicted quite scarce ability of active modes in both spectra, as is the case of RHF spectra in Fig. 10. Experimentally, both Raman [78] and IR [79-81] spectra of $C_{70}$ thin films is fine-structure rich similarly to those of the $C_{60}$ films. Comparing both experimental spectra with virtual ones in Fig. 10, a deep similarity of the former with the UHF ones is obvious. Once again, the feature evidences that the silent-mode-fine structure of vibrational spectra of $C_{60}$ and $C_{70}$ is the first-order effect caused by radical structure of the molecules.

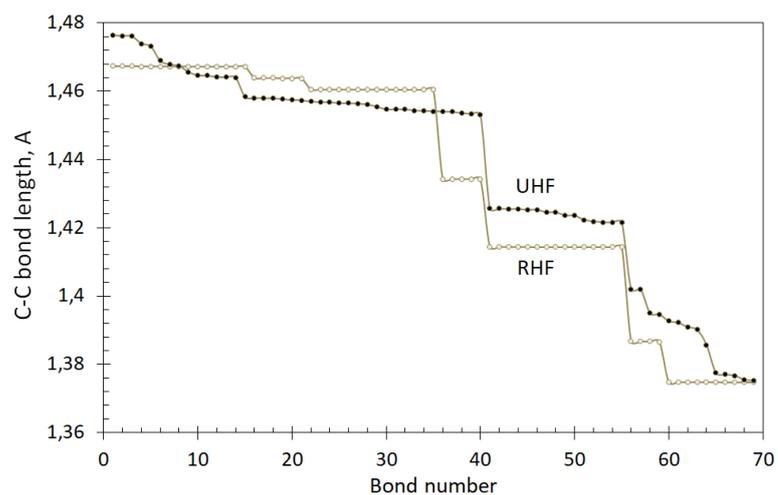

**Figure 9**. Correlation between C=C bonds of $C_{70}$ fullerene related to restricted and unrestricted HF calculations. Z→A distributions of $sp^2$C-C bonds.

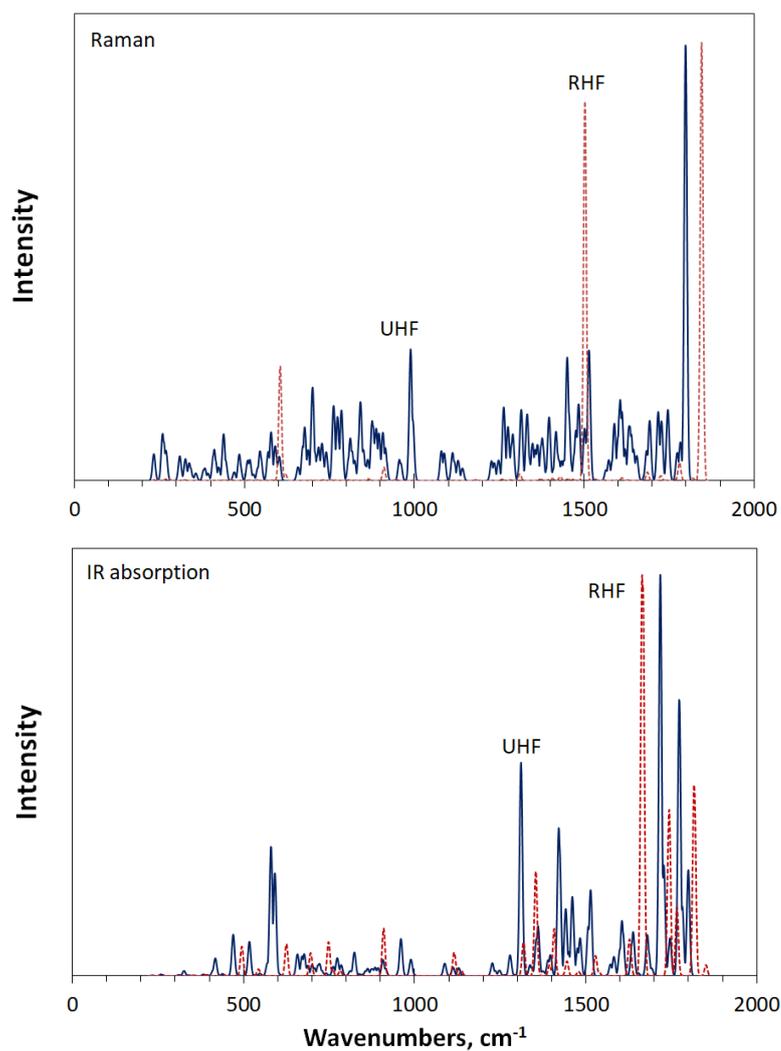

**Figure 10.** Raman and IR RHF (broken dark red) and UHF (blue) virtual spectra of fullerene $C_{70}$. Stick-bar data are convoluted with Gaussian bandwidth of 10 $cm^{-1}$. Intensities are reported in arbitrary units.

**Conclusion**

Unexpectedly for the authors, the UHF virtual vibrational spectrometer turned out to be effective, despite the fundamental shortcomings in the description of the force field of molecules. The computational program underlying the spectrometer does not allow correcting the force field, as a result of which it is impossible to avoid overestimated values of vibration frequencies. As a result, the performed testing of the spectrometer was limited to the general form of Raman and IR spectra in order to establish its suitability for reproducing the spectra of radicals and to determine the spectral characteristics indicating the radical nature of the studied molecules. The conclusion about the results obtained was made on the basis of a comparative analysis of the Raman and IR spectra produced with UHF virtual spectrum and their comparison with the available experimental data.

For the analysis, two polyacene and two fullerenes $C_{60}$ and $C_{70}$ were selected, at the beginning of the analysis there was benzene. It was found that the experimental spectra of naphthalene and pentacene are consistent with the UHF virtual spectra. In this case, if the Raman scans follow the main trend in the transformation of the benzene spectrum on going to polyacenes, then the UHF spectra reproduce the enhancement and development of the fine structure of the spectra, which corresponds to the experiment. In the case of fullerenes, this trend made it possible to establish the first order of the effect related to 'silent' modes in their spectra. The results obtained inspire confidence that the application of the UHF VVS to graphene molecules, the individual vibrational spectroscopy of which has not been studied, will reveal the main trends that determine their filling with active modes thus explaining the general shape of the spectra.

**Acknowledgements**

The publication has been prepared with the support of the "RUDN University Program 5-100".